\documentclass{ws-ijmpd}

\begin{document}

\markboth{Coelho, Lenzi, Malheiro, Marinho Jr. and Fiolhais}
{Investigation of the existence of hybrid stars using Nambu-Jona-Lasinio models}

%
\catchline{}{}{}{}{}
%

\title{INVESTIGATION OF THE EXISTENCE OF HYBRID STARS USING NAMBU-JONA-LASINIO MODELS}

\author{J. G. COELHO$^\ast$, C. H. LENZI$^\dagger$, M. MALHEIRO$^\ddag$, and R. M. MARINHO Jr.$^\S$}

\address{Departamento de F\'{i}sica, Instituto Tecnol\'{o}gico de Aeron\'{a}utica\\
 Pra\c{c}a Marechal Eduardo Gomes, 50, S\~{a}o Jos\'{e} dos Campos, S\~{a}o Paulo 12228-900, Brazil\\
$^\ast$jaziel@ita.br\\
$^\dagger$chlenzi@ita.br\\
$^\ddag$malheiro@ita.br\\
$^\S$marinho@ita.br}

\author{M. FIOLHAIS$^\odot$}

\address{Centro de F\'{i}sica Computacional, Departamento de F\'{i}sica, Universidade de Coimbra\\
Coimbra, P-3004-516, Portugal\\
$^\odot$tmanuel@teor.fis.uc.pt}
\maketitle

\begin{history}
\received{Day Month Year}
\revised{Day Month Year}
\comby{Managing Editor}
\end{history}

\begin{abstract}
We investigate the hadron-quark phase transition inside neutron
stars and obtain mass-radius relations for hybrid stars.
The equation of state for the quark phase using the standard NJL model
is too soft leading to an unstable star and suggesting a modification of the NJL
model by introducing a  momentum cutoff dependent on the chemical potential.
However, even in this approach, the instability remains. In order to remedy the instability
we suggest the introduction of a vector coupling in the NJL model, which makes the EoS stiffer,
reducing the instability. We conclude that the possible existence of
quark matter inside the stars require high densities, leading to very compact stars.
\end{abstract}

\keywords{quark matter; hybrid stars; Nambu-Jona-Lasinio.}

\section{Nambu-Jona-Lasinio model}	
A study of the stability of quark matter was carried on in Ref.\cite{jaziel} using the Nambu-Jona-Lasinio model in the SU(2) version with a repulsive
vector coupling. In that work the pressure and the energy per particle, as a function of the baryonic density, was analyzed. We also
discussed the influence of the vector interaction in the equation of state (EoS) and studied quark stars composed of pure quark matter
with two flavors. In the present work we consider the same NJL Lagrangian in the form\cite{klevansky,buballa}:
\begin{align}
{\cal {L}}=\bar{q} (i\gamma^\mu \partial_\mu - m)q +g_S[(\bar{q}q)^2+(\bar{q}i\gamma_5\overrightarrow{\tau}q)^2]
-g_V(\bar q \gamma^\mu q)^2,
\end{align}
where $q$ is a fermion field with $N_{\rm f}=2$ flavors and $N_{\rm c}$ colors. Except for the bare mass $m$, the Lagrangian is chirally symmetric
$\rm{SU(2)_L\times SU(2)_R}$. We included interaction terms in the scalar-isoscalar, pseudoscalar-isovector and vector-isoscalar channels.
The $g_S$ and $g_V$ are the scalar and vector couplings, respectively, and they are assumed to be
constants with dimensions MeV$^{-2}$.

Expanding $\bar{q}q$ and $\bar q \gamma^\mu q$ we can derive the mean field thermodynamic potential at temperature $T$ and chemical
potential $\mu$. We restrict ourselves to the Hartree approximation. The thermodynamic potential density, $\Omega$, depends on two parameters, namely
the dynamical fermion mass, $M$, and the renormalized quark chemical potential, $\mu_{\rm R}$, which are related to the
scalar, $\langle \bar{\psi} \psi \rangle$, and vector, $\langle \psi^\dagger \psi\rangle$, densities at the chemical
potential $\mu$ through\cite{jaziel,klevansky}:
\begin{equation}\label{gapmass}
M=m-2g_S\langle \bar{\psi} \psi \rangle,
\end{equation}
\begin{equation}\label{gappotencialquimico}
\mu_R=\mu-2g_V\langle \psi^\dagger \psi\rangle.
\end{equation}

These are the NJL gap equations for the dynamical mass and the renormalized chemical potential. The vacuum
contribution to the thermodynamic potential density is divergent and has to be regularized. This regularization is performed
by introducing a cutoff, $\Lambda$. Once we know the thermodynamic potential density, other thermodynamic quantities such
as the baryon number density, the energy density and the pressure, can be calculated in the standard way\cite{jaziel}.

\section{Hybrid Stars Phenomenology}
In the interior of astrophysical compact objects such as a neutron star, the density of matter can be several times the
nuclear matter saturation density. Calculations based on microscopic equations of state, which include only nucleons as degrees of freedom, show
that the central density of the most massive neutron stars is $\sim 7-10$ times the saturation density of nuclear matter, $\rho_0$. The precise point
where the transition from nuclear matter to the quark matter occurs, is not known and only estimates have been put forward
in the literature\cite{baldo}. We can assume a first order phase transition, as suggested by lattice calculations, and
use the mentioned NJL model to describe the quark phase and estimate the phase transition within the compact object\cite{Blaschke}. This approach
has been suggested by several authors and there is a vast literature in this topic, including calculations with other relativistic  models
such as the MIT and the chromodieletric model (CDM)\cite{cdm1,cdm2}. Neutron stars having deconfined quarks in their interior
are known as hybrid stars.

\subsection{Phase Transition}
According to the Gibbs criterium, two phases are in thermodynamical equilibrium when their baryonic chemical potentials, temperatures and
pressures are equal, corresponding, respectively, to chemical, thermal and mechanical equilibrium\cite{glendenning}:
\begin{equation}
 \mu_{\rm H}=\mu_{\rm Q}=\mu,
\end{equation}
\begin{equation}\label{temp}
 T_{\rm H}=T_{\rm Q}=T,
\end{equation}
and
\begin{equation}
 P_{\rm H}(\mu, T)=P_{\rm Q}(\mu, T)=P.\\
\end{equation}
Here, the indices H and Q refer to the hadron and quark phases, respectively. For the hadron phase we use the equation of state known
as GM1 (Glendenning and Moszkowski)\cite{gm1} and for the quark phase, we use the
mentioned NJL model with vector coupling. The GM1 parameterization describes the hadronic matter in the interior of the star and the NJL 
model the quark matter star core.

\begin{figure}
\centering
\begin{minipage}[b]{0.5\linewidth}
\includegraphics[width=\linewidth]{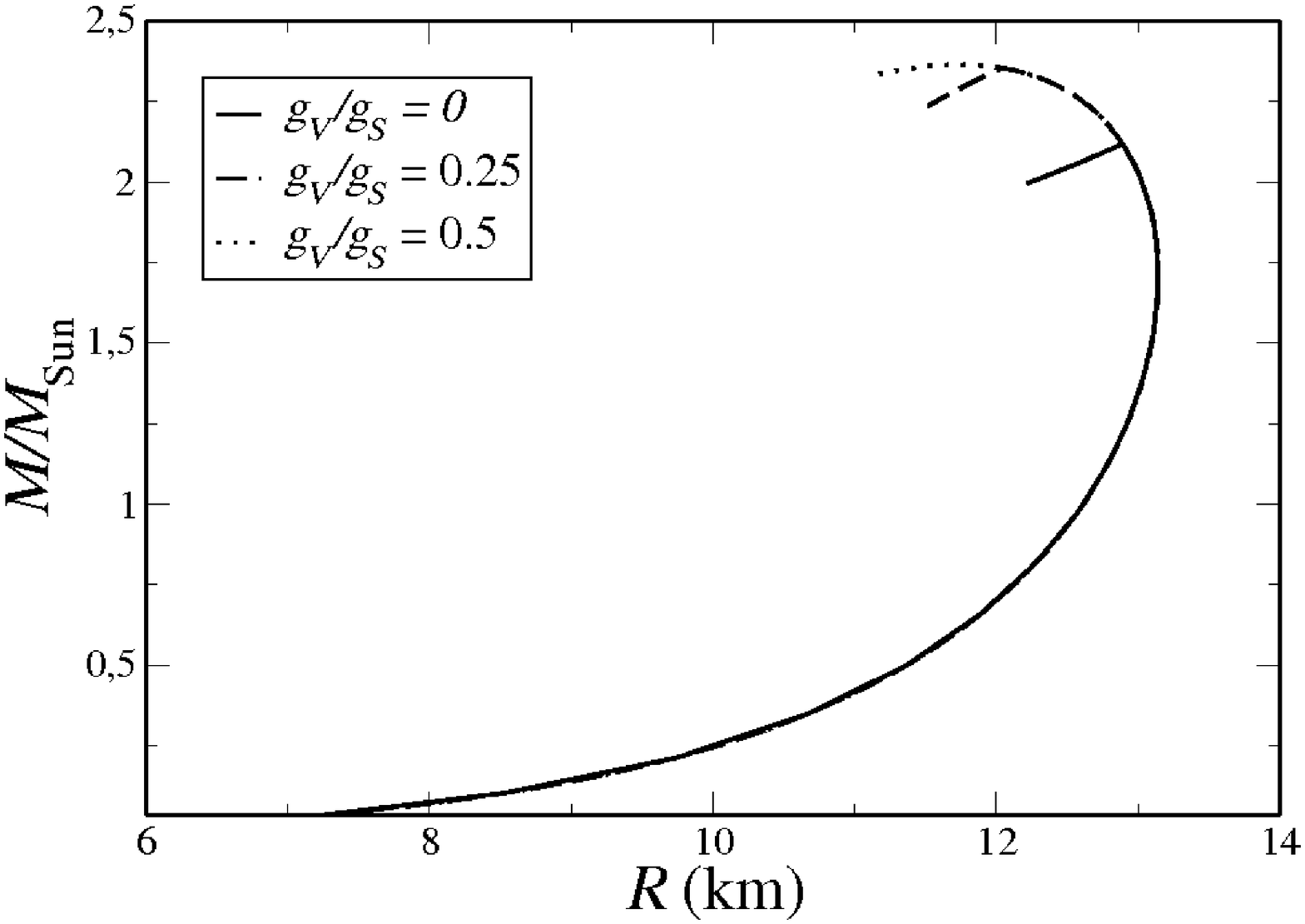}
\end{minipage} \hspace{-0.3cm}
\begin{minipage}[b]{0.5\linewidth}
\includegraphics[width=\linewidth]{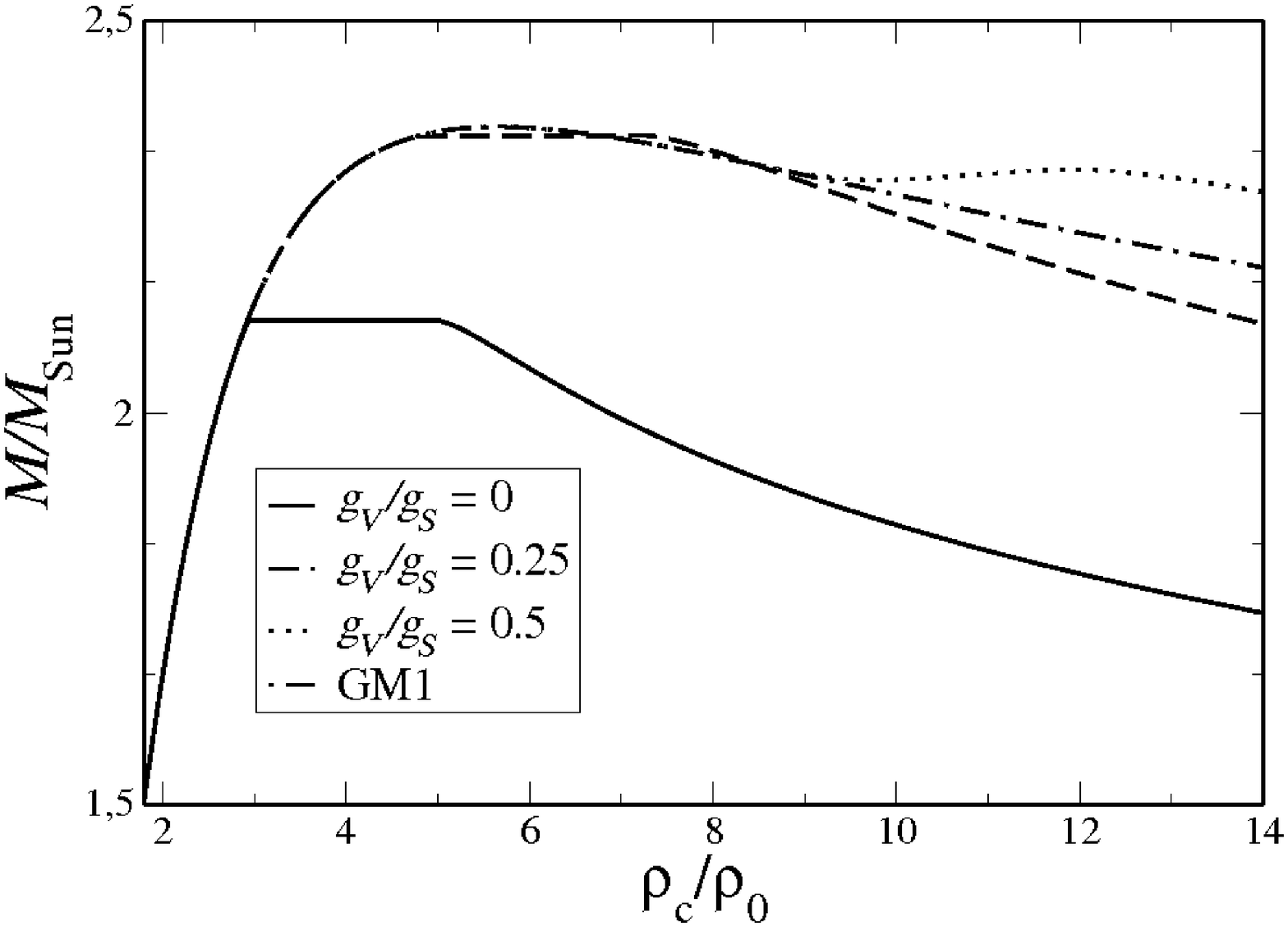}
\end{minipage}
\caption{\textit{Left panel}: Mass-radius diagram obtained with three
values for the vector coupling. \textit{Right panel}: Star mass plotted as a function of the
central density for different values of the vector
coupling.}\label{fig1}
\end{figure}

The EoS for quark matter obtained in Ref.\cite{jaziel} is used to solve the Tolman-Oppenheimer-Volkoff equations. These equations
are integrated from the center of the star, where $M(R =0)=0$ and $\varepsilon(R = 0)=\varepsilon_{\rm c}$, with $\varepsilon_{\rm c}$ the
central energy density, up to $r = R$, where the pressure drops to zero. For each EoS we obtain a unique relation between the star
mass and its central energy density\cite{glendenning,weber,taurines}.
Fig.~\ref{fig1}, illustrates the behavior of the mass of the neutron star as a function of the radius and of the central density for different
ratios $g_V/g_S$. We observe that the mass is characterized by a ``cusp'' which, according to Baldo \textit{et al.}\cite{baldo}, is
related to an instability. The plateau that appears in Fig.~\ref{fig1} (right) is a consequence of the
Gibbs criterium. The width of the plateau is related to the jump in the density at the onset of the quark matter in the interior of the star.
Baldo \textit{et al.}, based on the fact that the quark  EoS using the NJL model is too soft to yield a stable star with quark matter
in its interior, suggested a modified NJL model with a momentum cutoff  that depends on the chemical potential. However, even in this approach,
the instability remains\cite{baldo}. Therefore, we suggested the introduction of a vector coupling in the NJL model, making the EoS predicted by
this model more stiff, as discussed in Ref.\cite{jaziel}. Hence, including the vector term in the Lagrangian density, obtaining the EoS for a
quark phase and taking into account a hadron phase described  by the GM1 approach, we can investigate a possible phase transition,
finding, as Fig.~\ref{fig1} shows, that by increasing the vector coupling, the ``cusp'' that appears in the mass-radius graph, becomes more
tenuous, and the plateau that appears in the mass versus central density plot decreases from $g_V/g_S = 0 $ to $g_V/g_S =0.25$ and even disappears
at $g_V/g_S =0.5$. We do not show the effect of higher couplings, because, in that case, the phase transition starts at densities much higher than those
considered here ($\rm{\rho\leq 10\rho_0}$).

\section{Conclusions}
For a vector coupling such that $g_V/g_S = 0.5$ we observe that the quark phase shows up right before the second maximum in the plot of the mass
versus central density (the small central density range where $\frac{\partial (M/M_{\rm Sun})}{\partial (\rho_c/\rho_0)}> 0$).
We know that hadronic matter is stable up to the first maximum. The fact that there is  a second maximum,
means that in a small range of central densities, $\rho_{\rm c}/\rho_0 \sim 9$, the star is stable. The smallness of this region indicates 
that the possibility for stable hybrid stars is remote. We conclude that for the phase transition from nuclear to quark matter to take place for larger values of $g_V/g_S$ requires very high central densities, leading to more compact stars. Finally,  from  this study on hybrid
stars, we conclude that the use of the NJL model with vector coupling restricts the threshold value for the vector repulsion.

\section*{Acknowledgments}
This work was partially supported by FAPESP, CNPq and CAPES/FCT agreement 183/07.



\end{document}